\documentclass[12pt, english]{amsart}
\usepackage{babel}
\usepackage{geometry} % see geometry.pdf on how to lay out the page. There's lots.
\usepackage{graphicx}
\title{Porous Structure Design in Tissue Engineering Using Anisotropic Radial Basis Function}
\author{
	Ke, Liu (\texttt{keliu@uwm.edu})\\
	Ye, Guo\\
	Zeyun, Yu\\
	University of Wisconsin Milwaukee
}

\date{} % delete this line to display the current date

\begin{document}
%===========================================================
\begin{abstract}
Development of additive manufacturing in last decade greatly improves tissue engineering. During the manufacturing of porous scaffold, simplified but functionally equivalent models are getting focused for practically reasons. Scaffolds can be classified into regular porous scaffolds and irregular porous scaffolds. Several methodologies are developed to design these scaffolds. A novel method is proposed in this paper using anisotropic radial basis function (ARBF) interpolation. This is method uses geometric models such as volumetric meshes as input and proves to be flexible because geometric models are able to capture the characteristics of complex tissues easily. Moreover, this method is straightforward and easy to implement.

Keywords: additive manufacturing, tissue engineering, anisotropic radial basis function, geometric models
\end{abstract}

\maketitle

%===========================================================
\section{Introduction}
In order to improve biological tissues, tissue engineering (TE) which uses a scaffold to form new tissues for a medical purpose has a wide range of applications. Part of the applications in practice are repairing or replacing portion of or whole tissues and performing specific biochemical functions. Because of the inherent ability to produce customized porous scaffolds with different required architectures, the development of additive manufacturing (AM) techniques during last decade greatly improves tissue engineering. The latest ASTM standards defines AM as "a process of joining materials to make objects from three-dimensional (3D) model data, usually layer upon layer, as opposed to subtractive manufacturing methodologies" \cite{astm10}. More specifically, additive manufacturing starts from a 3D computer model and builds the final product by the addition of material, usually from a layer-by-layer fashion. This is a new manufacturing techniques comparing to conventional subtractive processes which removes material from a 3D block. Commercial AM techniques to fabricate scaffolds for tissue engineering applications include selective laser sintering (SLS), stereolithography (SLA), fused deposition modeling (FDM), precision extrusion deposition (PED) and 3D printing (3DP). Interested readers can refer to \cite{melchels12,bartolo09,peltola08} for details of these techniques.

Because native tissues are inherently heterogeneous and often have complex physiological architectures, literature, in practice, is primarily focused on the manufacturing of models which are simplified but functionally equivalent to the tissue to be repaired in terms of porosity and mechanical properties. Two types of porous scaffolds, namely regular porous scaffolds and irregular porous scaffolds, can be designed to achieve this goal. Numerous methodologies are proposed and categorized to fabricate these two types of scaffolds \cite{giannitelli14}. Table \ref{tab:scaffold_methods} categorizes these methods.

Periodic porous structures have a limitation that slight local modifications can affect the entire structure globally. Our proposed method provides a new implicit approach to generate porous tissue scaffold through volumetric mesh. Thus scaffold architectures can be adjusted by only modifying the mesh. Our proposed method has three major advantages over other implicit methods like TPMS-based ones. Firstly, local modifications of pore shape, size or distribution is achieved by changing local mesh accordingly, which is easy because there are only geometric changes in the mesh. Secondly, our method is flexible to simulate heterogeneities and discontinuities in natural tissue structures by using purposely-designed mesh as input. Depending on the features of tissue structure, meshes with different type (such as tetrahedron mesh and hexahedron mesh), size and density can be used to represent these characteristics. With different meshes as input, our method is able to build different tissue scaffolds with slight modifications in algorithms. Thirdly, unlike many implicit methods need post-actions like Boolean operations to get the final pieces built, the only post-action in our method is taking iso-surfaces, which is easier, faster and more flexible that user can get different scaffold architectures by taking different iso-values. In general, our method is superior in flexibility and easy to implement.

The rest of paper is organized as the following. Section \ref{sec:prior_works} summarizes progresses so far of methods categorized in Table \ref{tab:scaffold_methods}. Section \ref{subsec:rbf_interpolation} introduces porous scaffold reconstruction using conventional RBF interpolation. The proposed scaffold reconstruction is presented in Section \ref{subsec:arbf_interpolation}. Section \ref{subsec:algorithms} shows the overall algorithms of our proposed method. Finally, Section \ref{sec:results} shows some experimental results and discussions. Section \ref{sec:conclusions} concludes this paper.

%-.-.-.-.-.-.
\begin{table}
	\begin{tabular}[t]{l|c}
		\hline
		Type of scaffolds & Method \\
		\hline
		Regular porous scaffolds & CAD-based methods \\
		& Image-based methods \\
		& Implicit surface modeling (ISM) \\
		& Space-filling curves\\
		\hline
		Irregular porous scaffolds & An optimization method proposed by \cite{khoda11} \\
		& Stochastic methods using Voronoi models \cite{schroeder05,sogutlu07} \\
		& A hybrid Voronoi-spline method \cite{schaefer86} \\
		& Methods using volumetric meshes \cite{melchels11} \\
		\hline
	\end{tabular}
	\caption{Category of methods to design porous scaffolds in tissue engineering.}
	\label{tab:scaffold_methods}
\end{table}

%===========================================================
\section{Prior Works}\label{sec:prior_works}
CAD-based methods, such as constructive solid geometry (CSG) and boundary representation (B-Rep), are used to design regular porous scaffolds. CSG-based tools combine standard solid primitives (cylinders, spheres or cubes) through Boolean operations (e.g. intersection) to design and represent complex models. B-Rep tools describe the solid cell through its boundaries by a set of vertices, edges and loops without explicitly specify relations among them. So a preliminary check is required to verify there are no gaps or overlaps among the boundaries \cite{chiu06}. However, as objects become large or their internal architectures become more complex, their size increases dramatically hence it is hard or impossible to visualize and manipulate them. To overcome this limitation of most CAD-based tools, different solid cells with more bio-inspired features have been introduced \cite{sun05,bucklen08}.

Image-based methods combine imaging, image processing and free-form fabrication techniques to simplify scaffold design. Scaffolds can be described by 3D binary images (i.e. voxel values are Boolean and correspond to "solid" and "void"). Image-based methods produce scaffolds by taking the intersection of two 3D binary images, one representing the shape to be reproduced, and the other consisting of stacking of a binary unit cell. Empirically derived geometries are created in the unit cell with basic geometric shapes (cylinders, spheres) to represent regular pores within a scaffold. Randomly arranged pores can be obtained by the use of a random number generator to set voxel state. The topological optimization algorithms has been proved pivotal to obtain scaffolds in image-based methods \cite{hollister02, hollister05}.

Implicit surface modeling is highly flexible and describes scaffold architecture by a single mathematical equation, with freedom to introduce different pore shapes, pore size gradients. Recently, a large class of periodic minimal surfaces methods such as triply periodic minimal surfaces (TPMS) have become attractive for the design of biomorphic scaffold architectures. An early attempt of using TPMS-based method to control tissue fabrication is presented by \cite{rajagopalan06}. TPMSs like Schwartz's Primitive (type P), Schwartz's Diamond (type D) and Schoen's Gyroid (type G) are demonstrated their efficacy in high-precision fabrication of TPMS-based scaffolds \cite{seck10,melchels09,elomaa11}. However, all aforementioned works are limited to simple cubic or cylindrical outer shape. An improved method for constructing a pore network within an arbitrary complex anatomical model has been developed and successively optimized by Yoo et al. \cite{yoo11_1,yoo11_2}. In general, the TPMS methodologies, combining the advantages of both traditional CSG and image-based methods, are computationally efficient for modeling and fabrication of scaffolds.

Space-filling curves methodologies coupled to extrusion-based techniques. Such techniques consist of the extrusion of a micro-diameter polymeric filament terminating with a nozzle having an orifice diameter in the hundreds of microns range. The fabrication process involves the deposition of polymeric layers, which adhere to each other by heating temperature while retaining their shape. This process leads to regular repetition of identical pores. Thus these geometries have been named honeycomb-like patterns \cite{zein02}. More complex patterns can be obtained by changing the deposition angle between adjacent layers. An alternative approach to design scaffold is using fractal space-filling curves, which can be mathematically generated by starting with a simple pattern that grows through the recursive rules.

Periodic porous structures have several advantages. They are easier to model and their structural properties are possible to predicate. Their disadvantages lie in the difficulty of controlling the pore shape, size and distribution since slight modification of the unit cell will pose global changes to the entire structure. Moreover, current CAD tools are not suitable to reproduce the complex natural structures. In scaffold with variational porous architectures, discontinuities of deposition path planning are often found at the interface of two adjacent regions \cite{kalita03}. To design such scaffold architecture, Khoda et al. implemented an optimization method \cite{khoda11}. Stochastic and Voronoi models have been used to generate random pores in scaffold design as well. Heterogeneous pores distributed according to a given porosity level are generated by stochastic methods in scaffold design \cite{schroeder05,sogutlu07}. To overcome the limitation of only simple spheres can be used to represent pores, a hybrid Voronoi-spline representation combined with a random colloid-aggregation model is proposed \cite{schaefer86}. The proposed method has been extended to implement graded pore sizes and pore distributions \cite{kou12}. Volumetric mesh generators derived from finite element tools are used to create heterogeneous porosity within a solid model as well \cite{melchels11}.

%===========================================================
\section{Methods}\label{sec:methods}
%-------------------------------------------------------------------------
\subsection{Review of Radial Basis Function (RBF) Interpolation}\label{subsec:rbf_interpolation}
The conventional radial basis function interpolation is given by 
\begin{equation}\label{eqn:rbf_1}
f(\textbf{x})=\sum_{i=1}^N{w_i\phi(\|\textbf{x}-\textbf{x}_i\|)}
\end{equation}
where the interpolated function $f(\textbf{x})$ is represented as a weighted sum of $N$ radial basis functions $\phi(\cdot)$, each centered differently at $\textbf{x}_i$ and weighted by $w_i$. Let $f_j=f(\textbf{x}_j)$. By given conditions $f_j=\sum_{i=1}^N{w_i\phi(\|\textbf{x}_j-\textbf{x}_i\|)}$, the weights $w_i$ can be solved by
\begin{equation}
\begin{bmatrix}
\phi_{11} & \cdots & \phi_{1N} \\
\vdots & \ddots & \vdots \\
\phi_{N1} & \cdots & \phi_{NN}
\end{bmatrix}
\begin{bmatrix}
w_1 \\
\vdots \\
w_N
\end{bmatrix}
=
\begin{bmatrix}
f_1 \\
\vdots \\
f_N
\end{bmatrix}
\label{eqn:rbf_matrix}
\end{equation}
where $\phi_{ji}=\phi(\|\textbf{x}_j-\textbf{x}_i\|)$. Once the unknown weights $w_i$ are solved, the value at an arbitrary voxel in porous architecture can be calculated by 
\begin{equation}\label{eqn:rbf_2}
s(\textbf{x})=\sum_{i=1}^N{w_i\phi(\|\textbf{x}-\textbf{x}_i\|)}
\end{equation}
In conventional RBF, the distance between point $\textbf{x} \in \mathbb{R}^d$ and center $\textbf{x}_i \in \mathbb{R}^d$ is measured by Euclidean distance. Let $r=\|\textbf{x}-\textbf{x}_i\|$, commonly used radial basis functions include:\\
%\begin{table*}[b]
%	\begin{tabular}{c|c}
%		\hline
%		Types of basis functions & Definition \\
%		\hline
%		Gaussian & $\phi(r)=e^{-(cr)^2}$ \\
%		Multiquadrics (MQ) & $\phi(r)=\sqrt{r^2+c^2}$ \\
%		Inverse multiquadrics (IMQ) & $\phi(r)=\frac1{\sqrt{r^2+c^2}}$ \\
%		Thin plate spline (TPS) & $\phi(r)=r^2ln(r)$ \\
%		\hline
%	\end{tabular}
%\end{table*}
Gaussian: $\phi(r)=e^{-(cr)^2}$\\
Multiquadrics (MQ): $\phi(r)=\sqrt{r^2+c^2}$\\
Inverse MQ (IMQ): $\phi(r)=\frac1{\sqrt{r^2+c^2}}$\\
Thin Plate Spline (TPS): $\phi(r)=r^2ln(r)$\\
where $c$ is a shape parameter. Shape parameter plays a major role in improving accuracy of numerical solutions. In general, the optimal shape parameter depends on densities, distributions and function values at mesh nodes. Choosing shape parameter has been an active topic in approximation theory \cite{wang02}. Interested readers can refer to \cite{divo07,kosec08,sarler06,vertnik06,vertnik09} for more details.

\begin{figure}[t]
\centering
\includegraphics[scale=0.6]{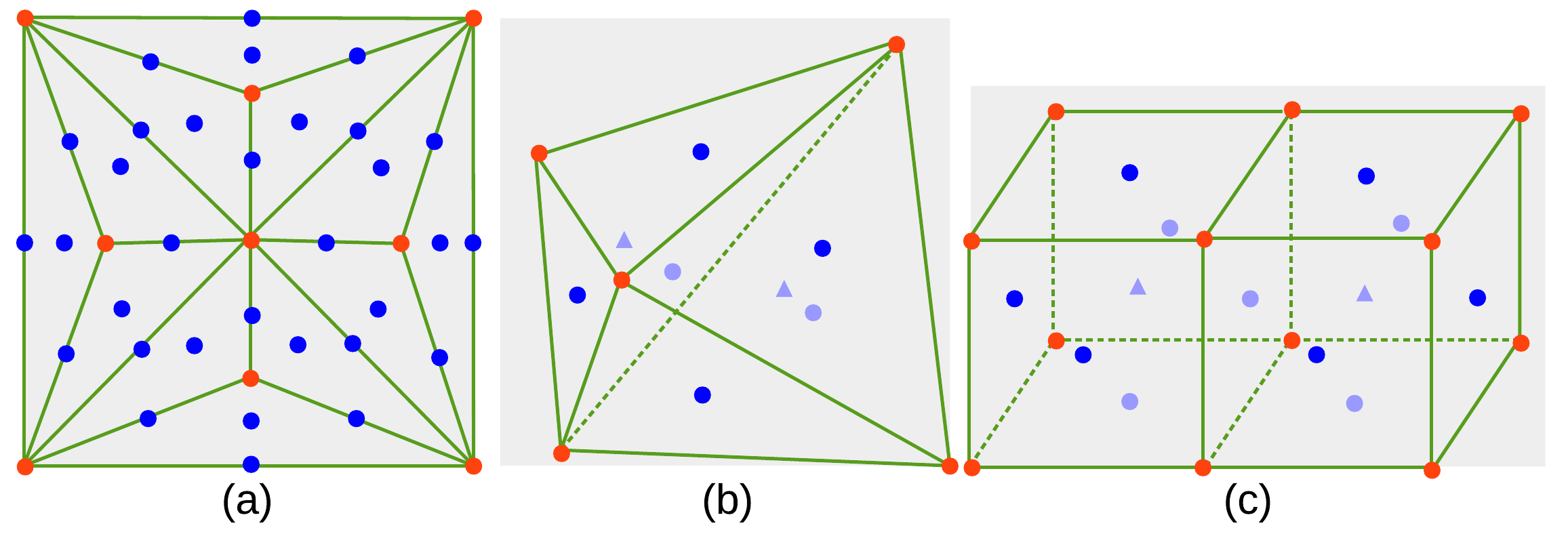}
\caption{Values assigned to mesh nodes. Red dots represent value of 1. Blue dots represent value of -1. In 3D meshes, interior dots are represented by lighter colors. (a) Sample 2D triangle mesh. Note both edge and triangle centers are given values -1. (b) Sample 3D tetrahedron mesh. Small triangle represent tetrahedron centers. (c) Sample 3D hexahedron mesh. Small triangle represent hexahedron centers.}
\label{fig:mesh_values}
\end{figure}

\begin{figure}[t]
\centering
\includegraphics[scale=0.85]{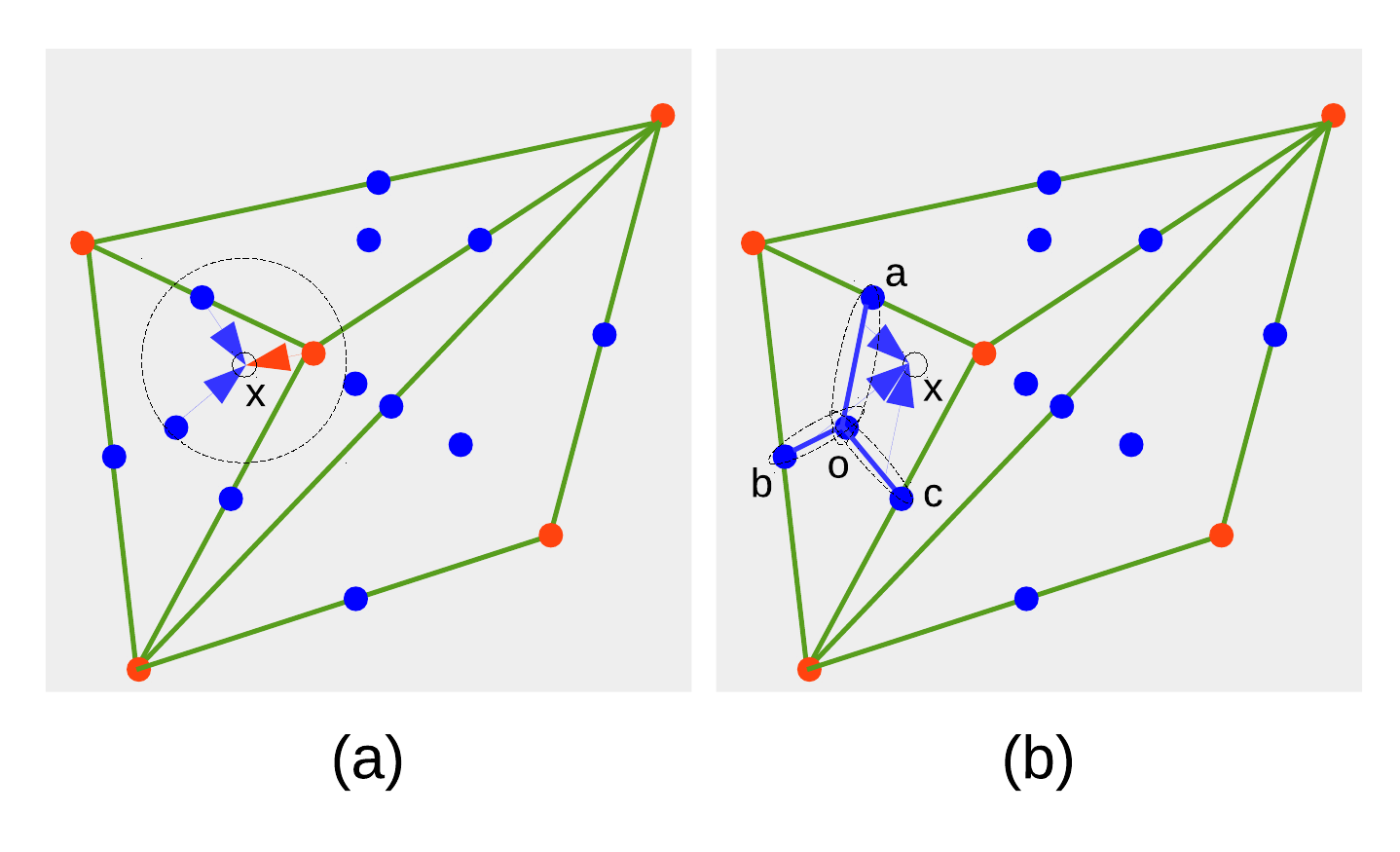}
\caption{2D interpolation schemes. x is the pixel to be interpolated. Dashed circle (for RBF) or ellipses (for anisotropic RBF) are support domains of underlying basis functions. (a) RBF interpolation. (b) Anisotropic RBF interpolation.}
\label{fig:interp_schemes}
\end{figure}

After interpolation, iso-surfaces are taken on the interpolated piece to create porous architecture. In detail, for both 2D and 3D meshes, mesh vertices are given values 1. For 2D meshes, edge centers and tile (triangle or quadrangle) centers are given values -1. For 3D meshes, face centers and sub-volume (tetrahedron or hexahedron) centers are given values -1. Fig. \ref{fig:mesh_values} (a-c) shows a sample 2D triangle mesh, a 3D tetrahedron mesh and a 3D hexahedron mesh with assigned values, respectively. For simplicity of drawing, 2D RBF interpolation scheme is illustrated by Fig. \ref{fig:interp_schemes} (a). 

Conventional RBF seems viable to construct porous scaffolds. However, since RBF is isotropic in the sense that only geometrical distance is considered, the support domains of underlying basis function always tend to be circular (in 2D) or spherical (in 3D), which sets an limitation to the customization of the internal architecture, especially for complex tissue scaffolds. Therefore, conventional RBF is only viable to generate simple architectures. Openings in desired scaffold architecture should start from sub-volume center and grows towards pores on structure surface. Given the condition that the face centers and sub-volume centers are assigned of value -1, the interpolated voxels should have values close to -1. Therefore, the internal opening directions have to be considered during interpolation and the shape of support domain should thus be anisotropic.
%-.-.-.-.-.-.

%-------------------------------------------------------------------------
\subsection{Anisotropic Radial Basis Function (ARBF) Interpolation}\label{subsec:arbf_interpolation}
The main difference between the isotropic and aniso-tropic RBFs is the definition of distance used. Given $N$ distinct line segments $L=\{l_j\}_{j=1,\ldots,N}$ the anisotropic radial basis function is defined by
\begin{equation}
\Phi_{L,j}(\cdot):=\phi(\|\cdot-l_j\|_L),
\end{equation}
where $\|\textbf{x}\|_L$ is defined as the distance between any arbitrary point and a line segment or the distance between two line segments.

\begin{figure}[t]
\centering
\includegraphics[scale=0.75]{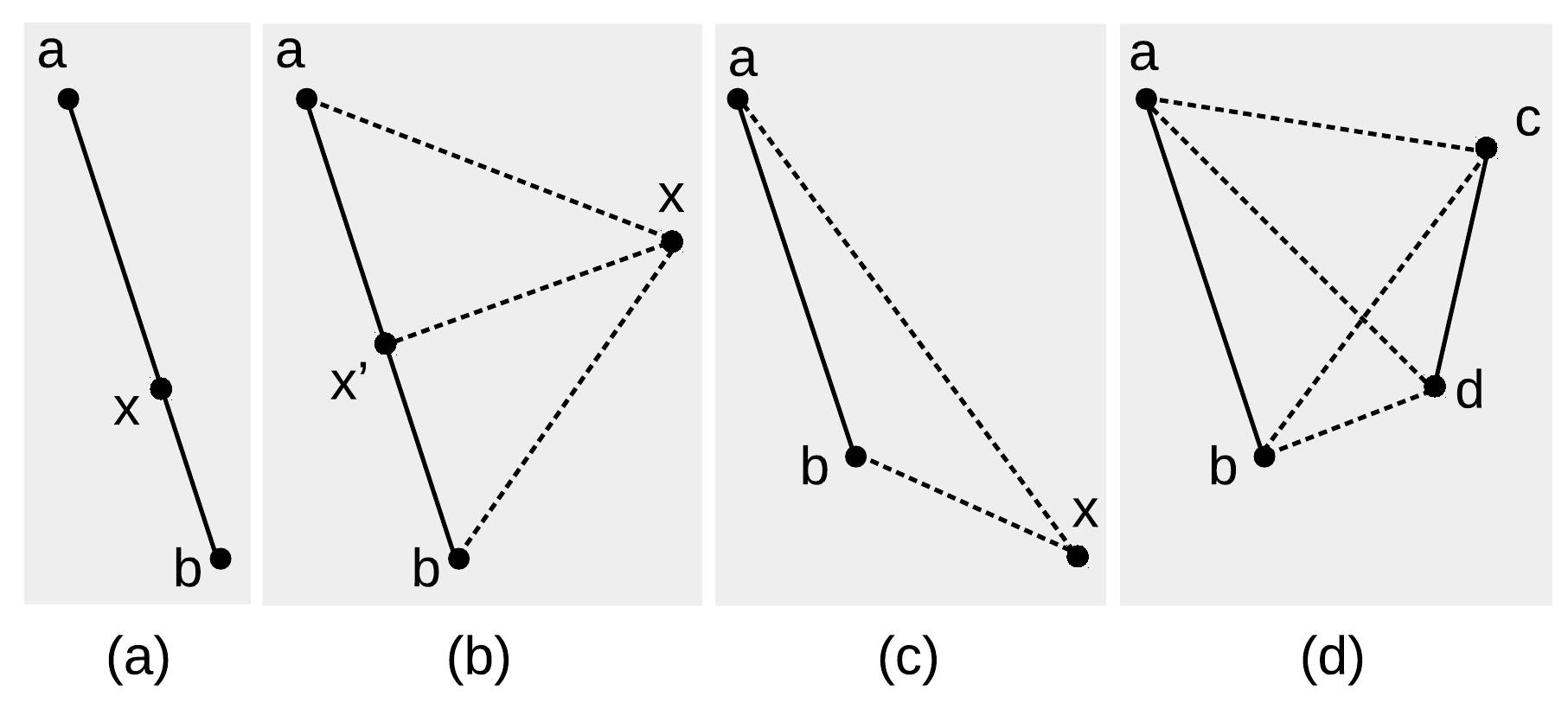}
\caption{Cases to calculate anisotropic distance. (a) Point $\textbf{x}$ is on line segment $(\textbf{a}, \textbf{b})$. (b) Point $\textbf{x}$ and line segment $(\textbf{a}, \textbf{b})$ form an acute triangle then distance is defined as the length of $\|\textbf{x}\textbf{x'}\|$. (c) Point $\textbf{x}$ and line segment $(\textbf{a}, \textbf{b})$ form an obtuse triangle. (d) Distance between line segment $(\textbf{a}, \textbf{b})$ and $(\textbf{c}, \textbf{d})$.}
\label{fig:aniso_distances}
\end{figure}

To calculate the distance between point $\textbf{x}$ and line segment $(\textbf{a}, \textbf{b})$, there are three cases to consider. For case 1, if point $\textbf{x}$ is on the line segment $(\textbf{a}, \textbf{b})$, distance is 0. For case 2, if point $\textbf{x}$, and $(\textbf{a}, \textbf{b})$ form an acute triangle, the distance is defined as the length of $\textbf{x}$'s projection to $(\textbf{a}, \textbf{b})$. For case 3, if $\textbf{x}$, and $(\textbf{a}, \textbf{b})$ form an obtuse triangle, the distance is defined as $min\{\|\textbf{x}\textbf{a}\|, \|\textbf{x}\textbf{b}\|\}$. Fig. \ref{fig:aniso_distances} (a-c) illustrates the three cases. 
The distance between any two arbitrary line segments $(\textbf{a}, \textbf{b})$ and $(\textbf{c}, \textbf{d})$ is defined as \\ $min\{\|\textbf{a}\textbf{c}\|, \|\textbf{a}\textbf{d}\|, \|\textbf{b}\textbf{c}\|, \|\textbf{b}\textbf{d}\|\}$. Fig. \ref{fig:aniso_distances} (d) illustrates this case. Table \ref{tab:types_distances} summaries these types of distances described above. 
Geometrically, the new distance changes support domain of basis function, from conventional circular or spherical shape to elliptical or hyper-elliptical shape. In other words, the new distance definition considers both magnitude and direction of line segment, which allows it to be a great tool to control the direction of internal structure. Customization of porous architecture can be achieved by measuring the distance between any arbitrary voxel and different line segments along with the distance between this voxel to other mesh nodes. Table \ref{tab:distance_definition} shows the three types of distance definition used during interpolation. In our experiments, the distances between sub-volume centers and tile centers are used. Fig. \ref{fig:interp_schemes} (b) illustrates this idea in simplified 2D triangular mesh. Comparing 2D and 3D meshes, sub-volume centers in 3D map to the tile centers in 2D, and tile centers in 3D map to the edge centers in 2D. As the figure shows, x is the pixel whose intensity is unknown. o is the center of a tile. a, b, c are the centers of three edges of a triangle. The distance between x and the three line segments (a, o), (b, o), (c, o) are calculated. Additionally, distances between the three line segments, namely, (a, o), (b, o) and (c, o) themselves are calculated as well. The final interpolation matrix is built on these distances along with the Euclidean distance between x to other mesh nodes.
Similar to isotropic RBF, with a modified distance, the ARBF interpolation problem becomes
\begin{equation}\label{eqn:arbf_1}
s'(\textbf{x})=\sum_{i=1}^N{w'_i \phi(\|\textbf{x}-\textbf{x}_i\|_L)}
\end{equation}
Please note that matrix in equation (\ref{eqn:rbf_matrix}) should also be updated accordingly with the new distance metric. Therefore, the new set of weights $w'_i$ would be different from the weights $w_i$ in isotropic interpolation. After interpolation, a proper iso-value is applied to the interpolated pieces to get the final result. Fig. \ref{fig:result_tetrahedron} and Fig. \ref{fig:result_hexahedron} show the ARBF interpolation results.

%-.-.-.-.-.-.

%-------------------------------------------------------------------------
\subsection{Algorithms}\label{subsec:algorithms}
The following algorithm illustrates major steps to build porous structure from triangular (2D), tetrahedron and hexahedron (3D) meshes. The primary step is ARBF interpolation which consists of three sub-steps. Firstly, values are assigned to mesh nodes as explained in Section \ref{subsec:rbf_interpolation} to build the matrix. Secondly, the weight coefficients are solved by using the new distance definition explained in Section \ref{subsec:arbf_interpolation}. Thirdly, the weights are used to interpolate the value at each voxel. After the piece is interpolated, an iso-surface is chosen and applied to get the final porous scaffold.
\\
\\
\noindent
{Algorithm: Porous Scaffold Construction}
\label{alg:construct_scaffold}
\begin{verbatim}
ConstructPorousScaffold()
{
    loadMesh(); // read mesh
    ARBFInterpolation(); // do ARBF interpolation
    marchCube(); // take iso-surface
    exportResult(); // print result
}

ARBFInterpolation()
{
    assignMeshValues(); // assign mesh nodes and build matrix
    solveCoefficients(); // solve unknown weights
    applyCoefficientsToInterpolation(); // do the actual interpolation
}
\end{verbatim}

%===========================================================
\section{Results and Discussion}\label{sec:results}
\begin{figure}[t]
\centering
\includegraphics[scale=0.7]{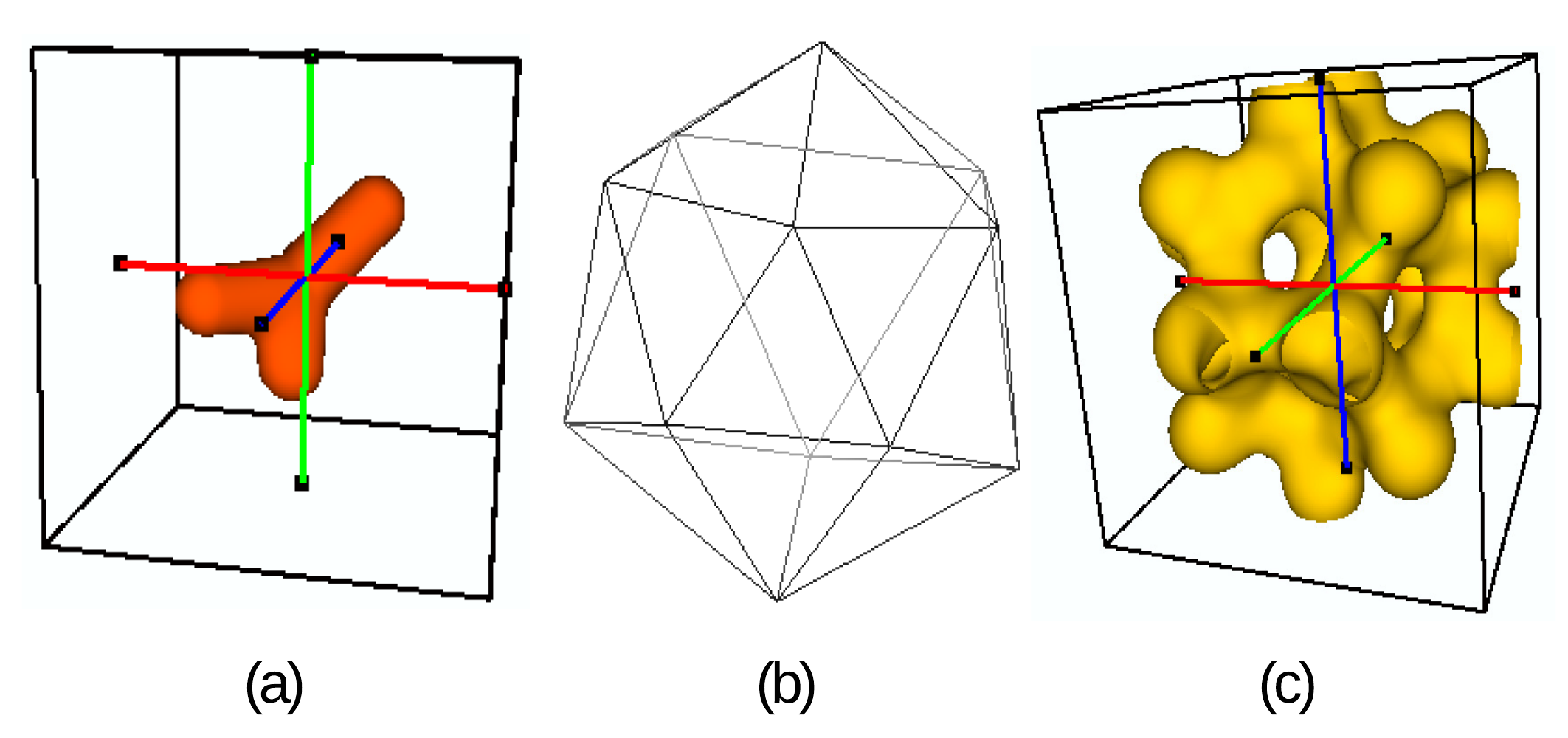}
\caption{Results based on tetrahedron meshes. (a) Scaffold based on single tetrahedron. (b) Input icosahedron mesh (formed by 20 tetrahedrons). (c) Scaffold using (b) as input.}
\label{fig:result_tetrahedron}
\end{figure}

\begin{figure}[t]
\centering
\includegraphics[scale=0.85]{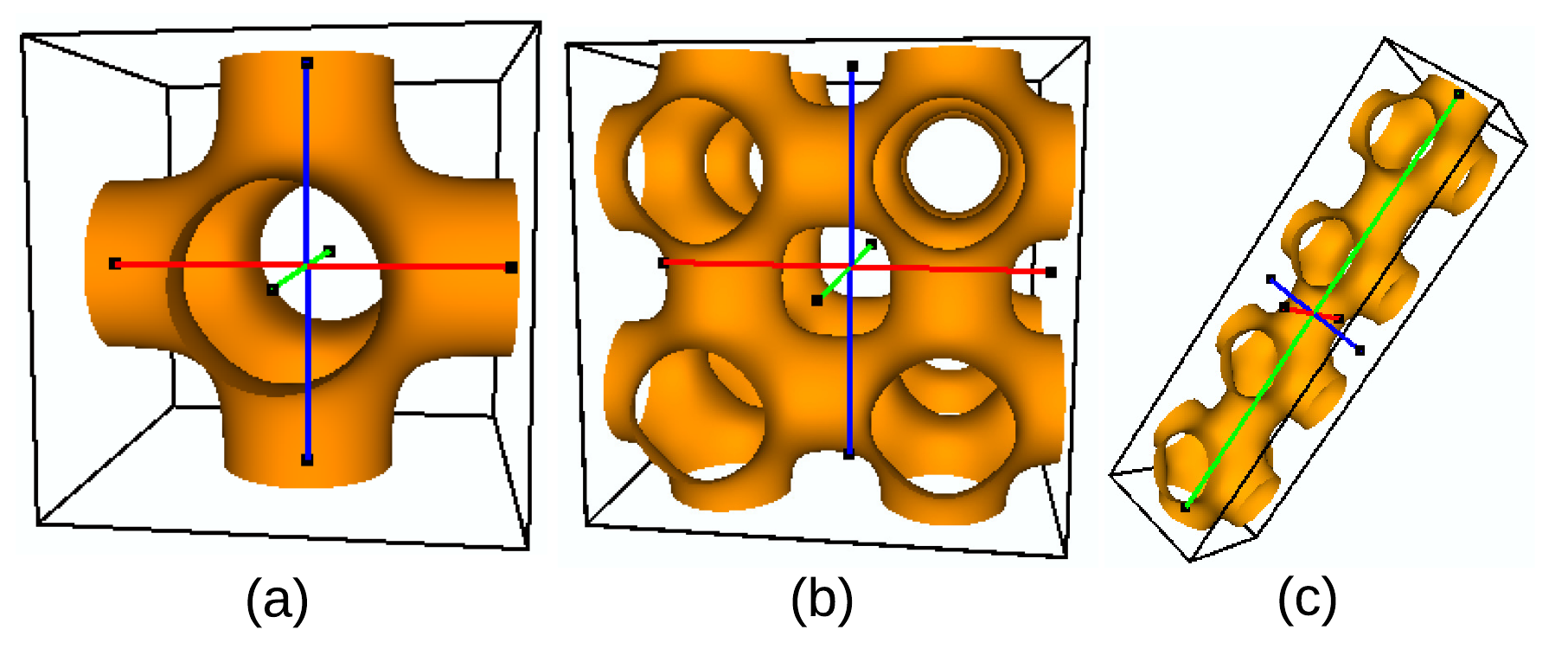}
\caption{Results based on hexahedron meshes. (a) Scaffold based on single hexahedron. (b) Scaffold based on 8 hexahedrons arranged to form a large cube. (c) Scaffold  based on 4 hexahedrons arranged to form a rod shape.}
\label{fig:result_hexahedron}
\end{figure}

\begin{figure}[t]
\centering
\includegraphics[scale=0.5]{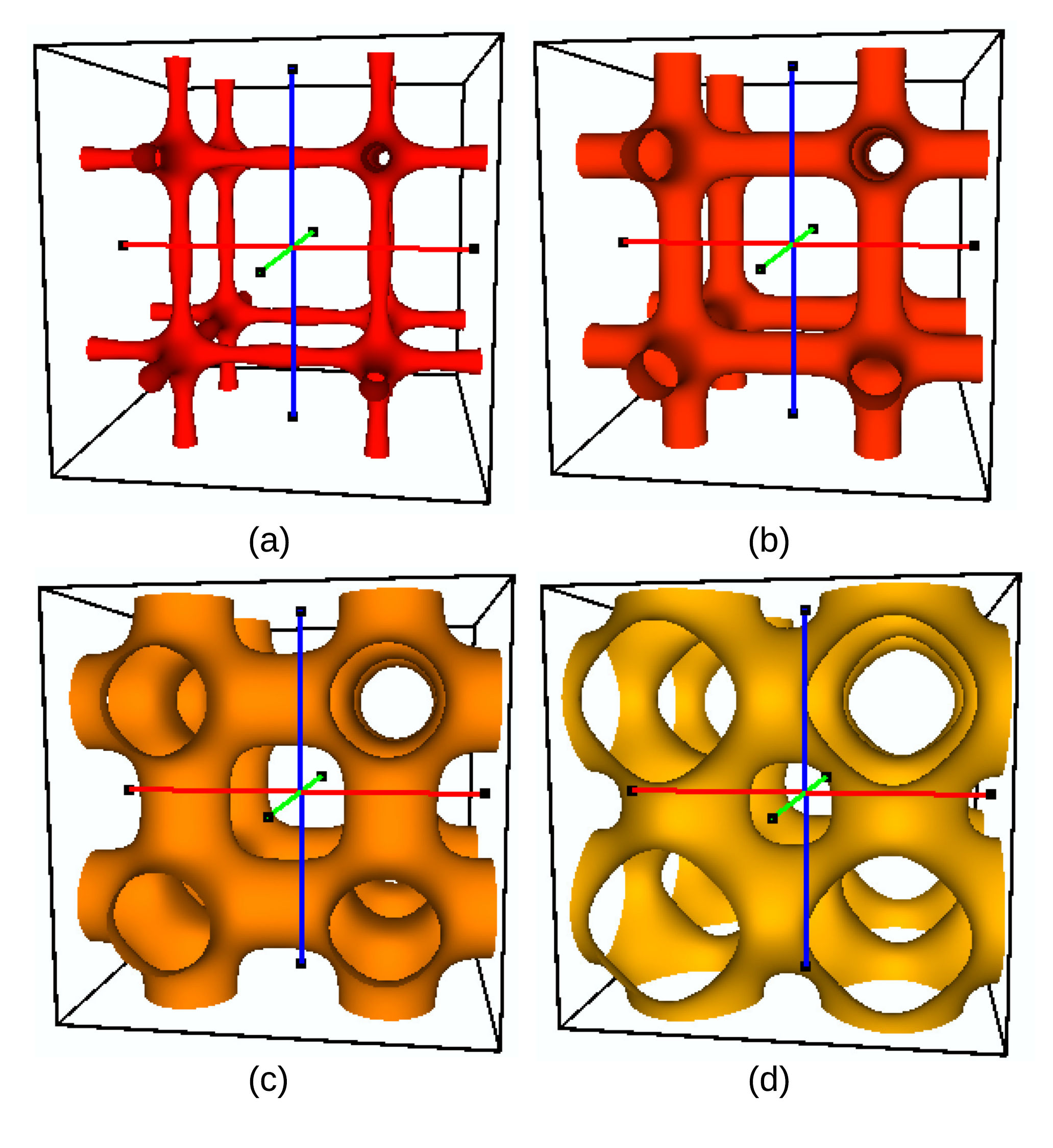}
\caption{Results taken by different iso-values. (a) - (d) are results taken by increasing iso-values.}
\label{fig:result_isosurfaces}
\end{figure}

\begin{figure}[t]
\centering
\includegraphics[scale=0.7]{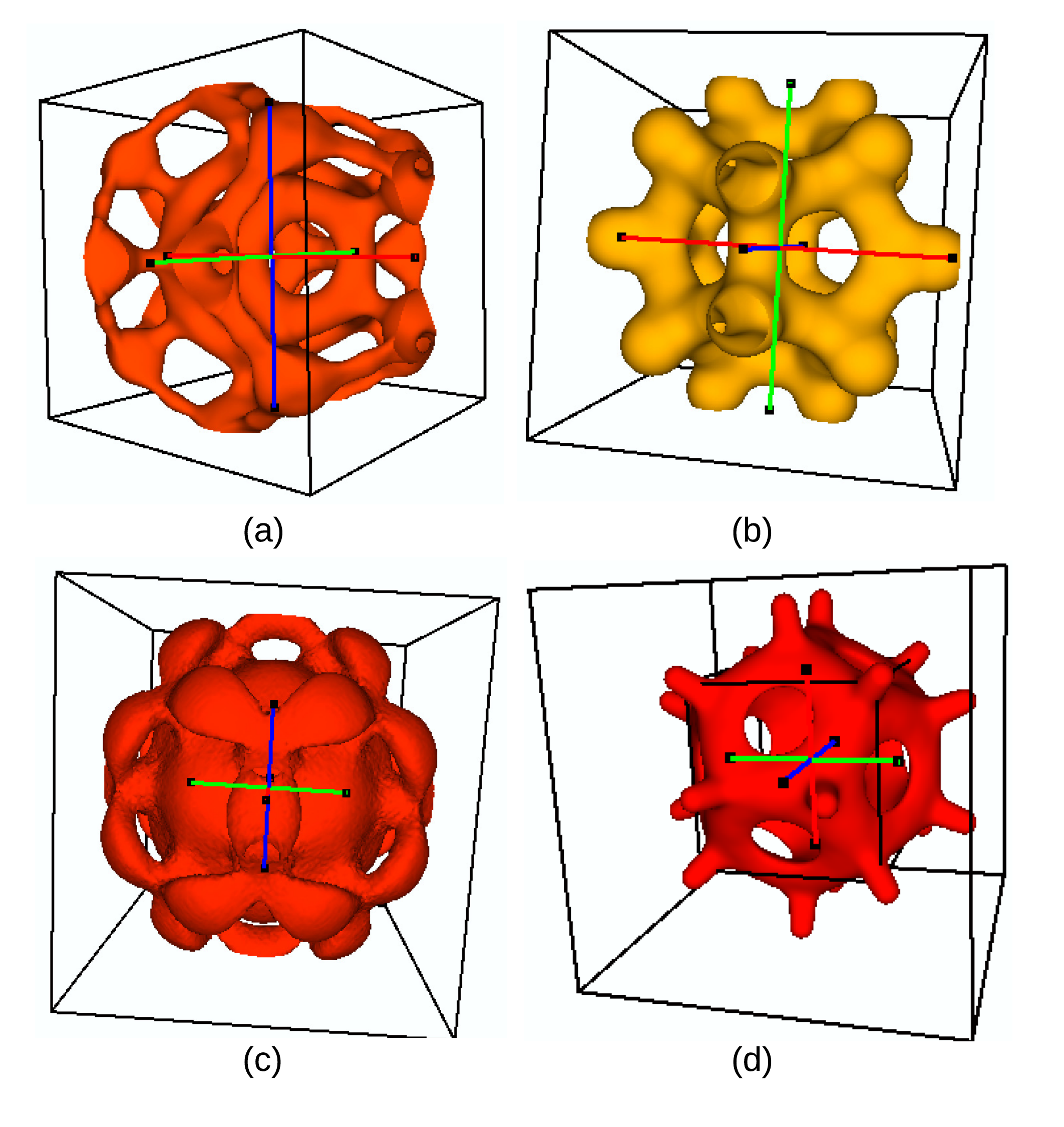}
\caption{Results obtained by using different basis functions in ARBF interpolation. Shape parameter is 0.1 for all results. (a) Result interpolated by multiquadrics (MQ) basis. b) Result interpolated by inverse multiquadrics (IMQ) basis. (c) Result interpolated by Gaussian basis. (d) Result interpolated by thin plate spline (TPS) basis.}
\label{fig:result_basis_funcs}
\end{figure}

\begin{figure}[t]
\centering
\includegraphics[scale=0.7]{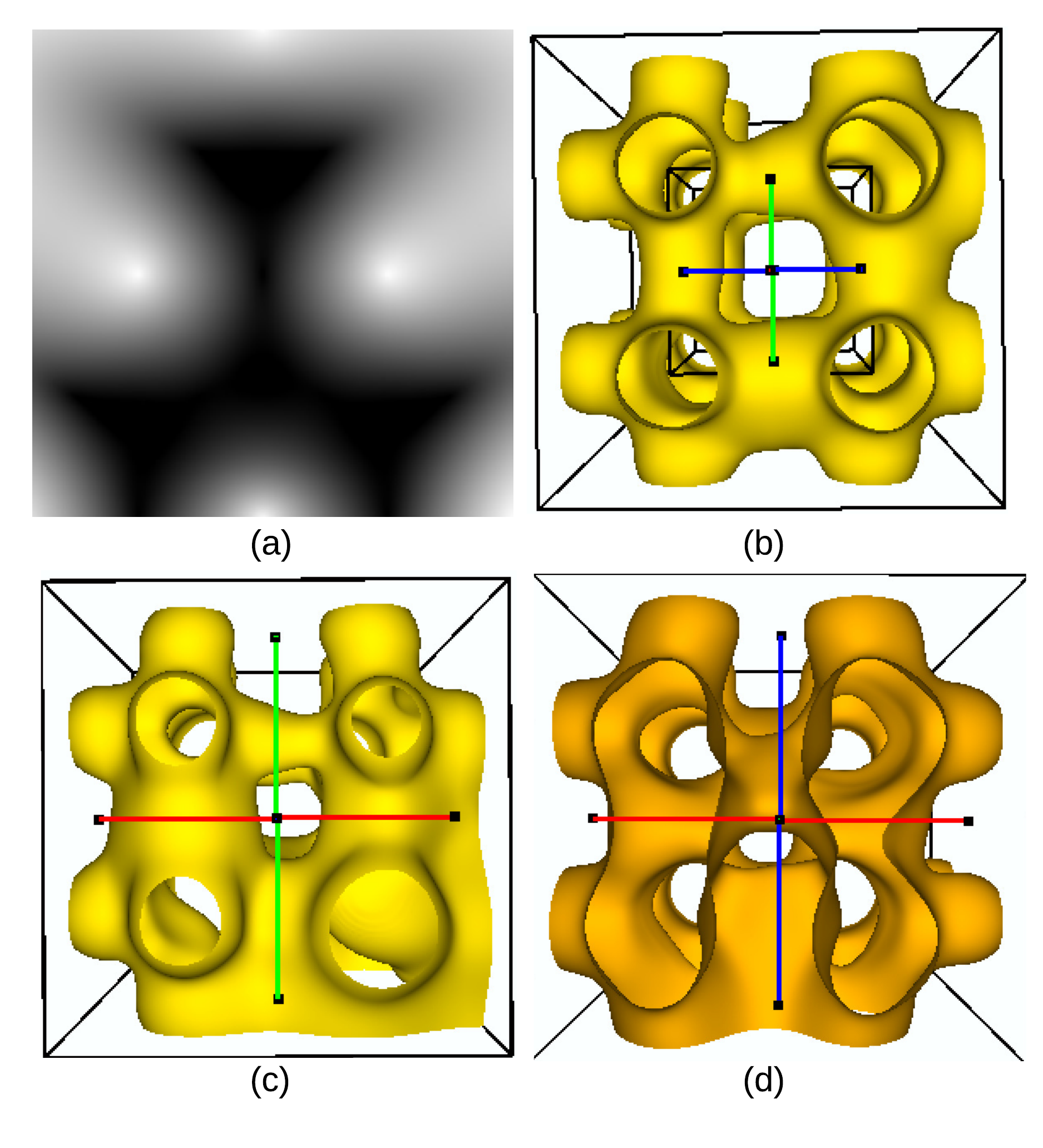}
\caption{Some experimental results. (a) Result interpolated by isotropic RBF and based on a 2D triangle mesh. (b-d) Results based on hexahedron meshes with disturbance.}
\label{fig:result_experimental}
\end{figure}

\begin{figure}[t]
\centering
\includegraphics[scale=0.75]{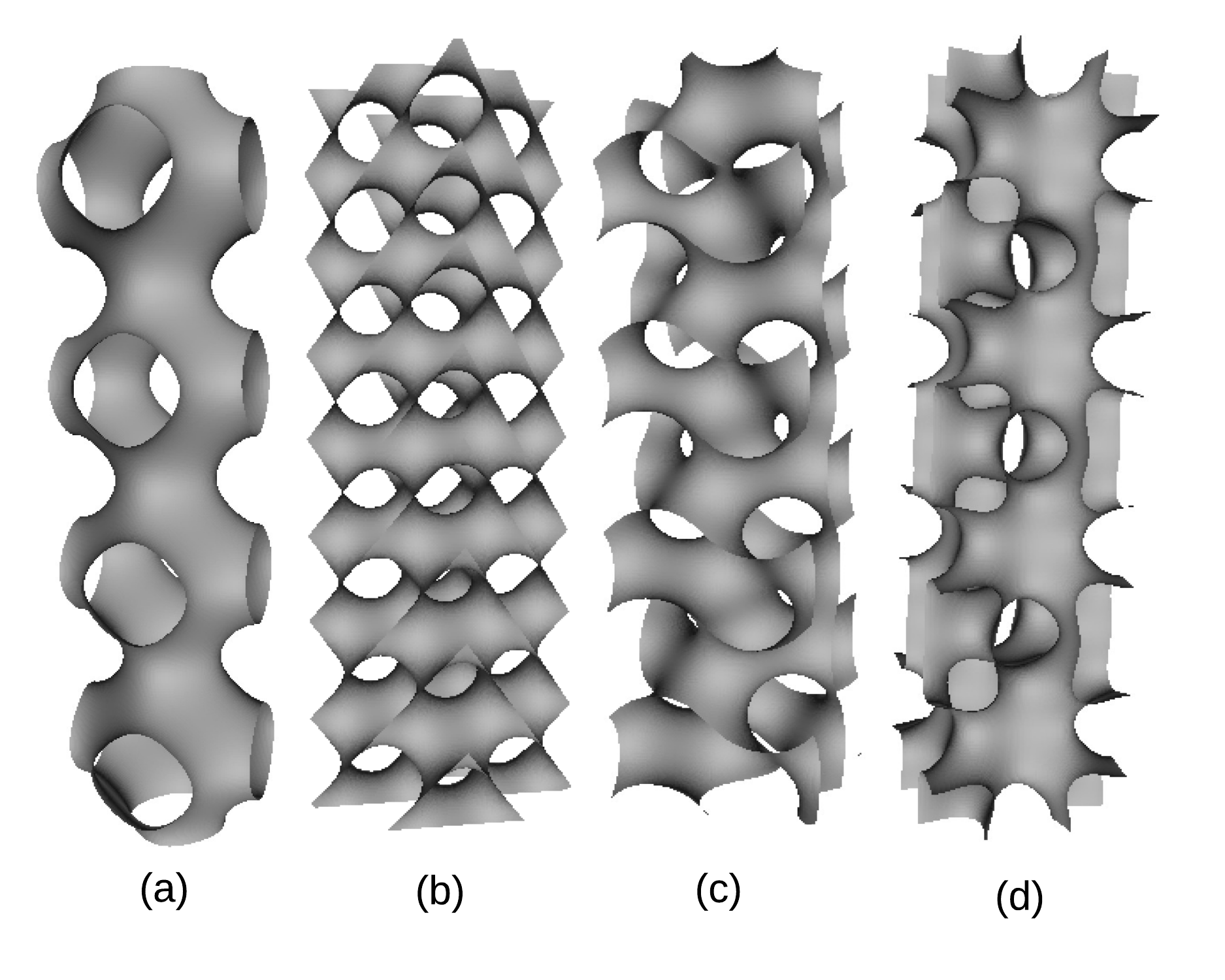}
\caption{Porous structure obtained by TPMS method. (a) Structure obtained by P-type function. (b) Structure obtained by D-type function. (c) Structure obtained by G-type function. (d) Structure obtained by IWP-type function.}
\label{fig:result_tpms}
\end{figure}

This section includes results of scaffold shapes using different types of meshes and shows several experimental results. Fig. \ref{fig:result_tetrahedron} (a) shows a scaffold obtained from single tetrahedron. The opening grows from tetrahedron center toward the four triangle centers. \ref{fig:result_tetrahedron} (b) illustrates an icosahedron mesh comprised of 20 tetrahedrons. \ref{fig:result_tetrahedron} (c) shows the scaffold interpolated from \ref{fig:result_tetrahedron} (b). Fig. \ref{fig:result_hexahedron} (a) shows a scaffold obtained from single hexahedron. Fig. \ref{fig:result_hexahedron} (b) shows a scaffold obtained from a block-shape hexahedron mesh which is comprised of 8 smaller hexahedrons. Fig. \ref{fig:result_hexahedron} (c) shows a scaffold obtained from a rod-shape hexahedron mesh which is comprised of 4 smaller hexahedrons. Input meshes for these results are regular so the final scaffolds are regular as well. The results are interpolated by ARBF interpolation introduced above using inverse multiquadrics (IMQ) as basis functions and shape parameter of value 0.1. After interpolation, a proper iso-value is applied to the interpolated piece to show the final porous structure. Because the input meshes are regular, the output structures tend to be regular as well. Iso-values can be adjusted to get structure of different size of pores. Results of different iso-values are included in Fig. \ref{fig:result_isosurfaces}. As the figure shows, pore sizes increase as the iso-values increases. So after interpolation, choosing a proper iso-value is required to obtain the desired porous structure. Besides iso-values, basis function can also affect the porous architecture in terms of mostly the size and shape of pores. Because the shape of basis functions are different, when the final piece is interpolated, different voxels are included in their support domains. Interpolated results using different basis functions are included in Fig. \ref{fig:result_basis_funcs}. To compare isotropic RBF interpolation and anisotropic RBF interpolation, a 2D isotropic RBF interpolated result (Fig. \ref{fig:result_experimental} (a)) is also included. As explained before, because basis functions in isotropic RBF have circular support domains, the circle-shape artifacts can be seen in the result. Additionally, interpolation results based on disturbed meshes are also investigated and shown in Fig. \ref{fig:result_experimental} (b-d). These disturbed meshes are generated by moving some mesh vertices in random direction. As the results shown, disturbed porous structures can be obtained from disturbed meshes. Therefore, adding disturbance is a viable way to construct porous structure with randomness, which is very useful to construct tissues with complex physiological architectures. Finally, as a contrast, TPMS results are also included in Fig. \ref{fig:result_tpms}.

%-.-.-.-.-.-.

%===========================================================
\section{Conclusions}\label{sec:conclusions}In practical, constructed scaffold may be very complicated to simulate complex physiological tissue architecture in terms of equivalent internal connectivity and mass transportation. Periodic porous structure cannot meet this challenge satisfactorily. Our proposed method uses volumetric mesh as input which can be very complex ¨C thanks to modern mesh modeling techniques. Thus, using complex mesh as input, our proposed method is able to construct complicated porous scaffold structures to meet this challenge. Moreover, modifications to the final structure can be achieved by common mesh operations or changing iso-value, which makes our method very flexible comparing to other period porous manufacturing techniques (like implicit surface methods). Finally, implementation of our proposed method is easy and computational cost is low because the core algorithm of interpolation is calculating distances and solving unknown weights. There are a lot of mature and fast linear algebra libraries available to use.
In the future, this method will be tested with adaptive meshes which represent porous architectures with heterogeneous and discontinuous structures. Moreover, a criteria to measure final scaffold will be developed to further test the efficacy and efficiency of this method.

%===========================================================

\end{document}